\PassOptionsToPackage{unicode}{hyperref}
\RequirePackage{textgreek}
\documentclass[a4paper,11pt]{article}

\usepackage{pos}

\usepackage{amsmath}

\usepackage{graphicx}
\usepackage{slashed}
\usepackage{booktabs,tabulary}
\usepackage{placeins}   

\newcommand{\kk}{\mathbf{k}}

\newcommand{\nn}{\mathbf{n}}
\newcommand{\mpi}{M_\pi}
\newcommand{\mN}{M_N}
\newcommand{\spiN}{\sigma_{\pi N}}

\newcommand{\MeV}{\,\text{MeV}}

\allowdisplaybreaks
\begin{document}

\title{The pion-nucleon sigma term from Lattice QCD}

\ShortTitle{The nucleon sigma term from Lattice QCD}

\author*[a]{Rajan Gupta}

\author[a]{Tanmoy Bhattacharya}

\author[b]{Martin Hoferichter}

\author[a]{Emanuele Mereghetti}

\author[a,c]{Sungwoo Park}

\author[d]{Boram Yoon}

\affiliation[a]{Los
Alamos National Laboratory, Theoretical Division T-2, Los Alamos, NM 87545, USA}

\affiliation[b]{Albert Einstein Center for Fundamental Physics, Institute for Theoretical Physics, University of Bern, Sidlerstrasse 5, 3012 Bern, Switzerland}

\affiliation[c]{Jefferson Lab, 12000 Jefferson Avenue, Newport News, Virginia 23606, USA}

\affiliation[d]{Los Alamos National Laboratory, Computer Computational and Statistical Sciences Division, CCS-7, Los
Alamos, NM 87545, USA}
      
\emailAdd{rajan@lanl.gov}
      
\abstract{We summarize recent evidence, both from lattice QCD and chiral perturbation theory, that suggests that larger-than-expected excited-state contamination could be the reason for the tension between phenomenological determinations and previous direct lattice-QCD calculations of the pion--nucleon sigma term $\spiN$. In addition, we extend the $\chi$PT analysis by calculating the corrections due to including the $\Delta(1232)$ resonance as an explicit degree of freedom. This correction is found to be small, thereby  corroborating the excited-state effects found in the $\Delta$-less calculation and the result for  $\spiN$.  
 }

\FullConference{The 10th International Workshop on Chiral Dynamics\\
		15--19 November 2021\\
		Beijing, China}

		\maketitle
		
\section{Introduction}

\begin{figure}[t]
\begin{center}
     \includegraphics[width=0.75\linewidth]{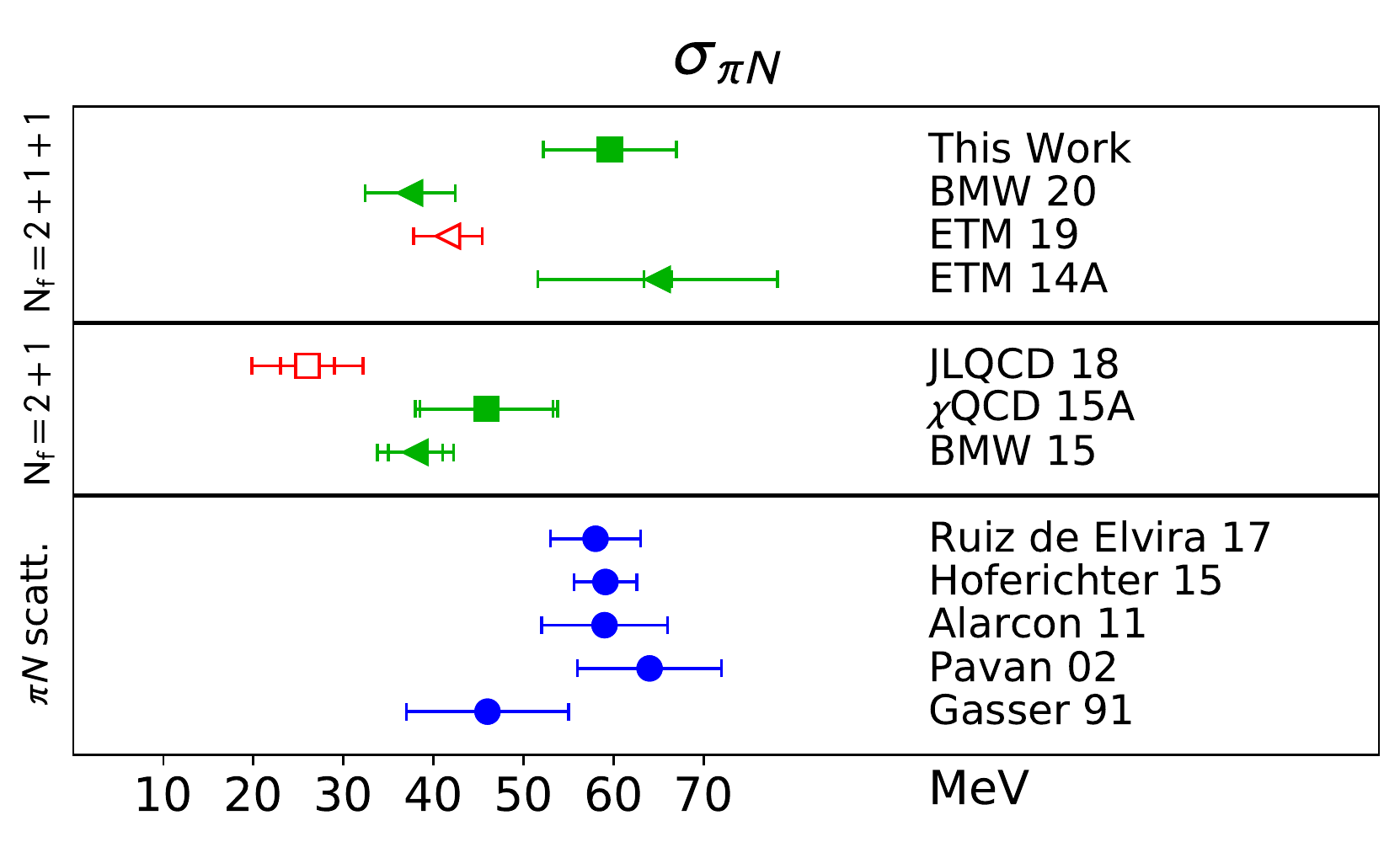}
\end{center}
\caption{Results for $\spiN = { m}_{ud}
  g_S^{u+d}$ from 2+1- and 2+1+1-flavor lattice calculations. 
  The BMW 20 result from 1+1+1+1-flavor lattices is listed along with 2+1+1-flavor calculations for brevity: the difference is expected to be insignificant. 
  Calculations in the
  direct approach are indicated by squares and the Feynman--Hellmann
  method by triangles.
The references from which lattice results have been taken are: 
JLQCD~18~\cite{Yamanaka:2018uud}, 
$\chi$QCD~15A~\cite{Yang:2015uis}, 
BMW~15~\cite{Durr:2015dna}, 
ETM~14A~\cite{Alexandrou:2014sha},
ETM 19~\cite{Alexandrou:2019brg}, and
BMW 20~\cite{Borsanyi:2020bpd}. 
  Phenomenological determinations from $\pi N$ scattering data (blue filled circles) are from 
  Gasser~91~\cite{Gasser:1990ce},
  Pavan~02~\cite{Pavan:2001wz},
  Alarcon 11~\cite{Alarcon:2011zs},
  Hoferichter 15~\cite{Hoferichter:2015dsa}, and 
  Ruiz de Elvira 17~\cite{RuizdeElvira:2017stg}. Figure reproduced from Ref.~\cite{Gupta:2021ahb}.}
\label{fig:FLAG}
\end{figure}

In Ref.~\cite{Gupta:2021ahb}, we proposed a possible resolution to a persistent tension between lattice-QCD and phenomenological determinations of the pion--nucleon $\sigma$-term $\spiN$. This quantity describes the coupling of the nucleon to an isosymmetric scalar current comprised of the two lightest quarks and appears prominently in searches for physics beyond the Standard Model whenever scalar currents play a role, e.g., in direct-detection searches for  dark
matter~\cite{Bottino:1999ei,Bottino:2001dj,Ellis:2008hf,Crivellin:2013ipa,Hoferichter:2017olk},
lepton flavor violation in $\mu\to e$ conversion in
nuclei~\cite{Cirigliano:2009bz,Crivellin:2014cta}, and electric dipole
moments~\cite{Engel:2013lsa,deVries:2015gea,deVries:2016jox,Yamanaka:2017mef}. Even though there is no scalar probe in the Standard Model, $\spiN$ can still be extracted from data on pion--nucleon ($\pi N$) scattering via the Chang--Dashen low-energy theorem~\cite{Cheng:1970mx,Brown:1971pn}. Such determinations have a long history~\cite{Gasser:1990ce,Pavan:2001wz}, with all recent determinations converging on a value around $60\MeV$, irrespective of whether the $\pi N$ input is taken from data on pionic atoms~\cite{Strauch:2010vu,Hennebach:2014lsa,Hirtl:2021zqf,Baru:2010xn,Baru:2011bw,Alarcon:2011zs,Hoferichter:2015dsa} or low-energy $\pi N$ cross sections~\cite{RuizdeElvira:2017stg}. In contrast, most lattice calculations~\cite{Durr:2015dna,Yang:2015uis,Yamanaka:2018uud,Alexandrou:2019brg,Borsanyi:2020bpd} (with the exception of Ref.~\cite{Alexandrou:2014sha}) prefer a value as low as $40\MeV$, producing the tension~\cite{Hoferichter:2016ocj} summarized in Fig.~\ref{fig:FLAG} and in the FLAG 2021 report~\cite{Aoki:2021kgd}.
These lattice calculations can be grouped into those applying the Feynman--Hellmann theorem, in which case the quark-mass derivative of the nucleon mass needs to be controlled very accurately, and via the direct calculation of the three-point function.  Reference~\cite{Gupta:2021ahb} provided lattice evidence that the mismatch between the direct method and phenomenology can be explained by larger-than-expected multihadron excited-state contamination (ESC). Motivation for such ESC was also presented using chiral perturbation theory ($\chi$PT). Here, after a short review of the lattice-immanent arguments given in Sec.~\ref{sec:lattice}, we extend the $\chi$PT calculation of ESC for the $\Delta$-less case presented in Ref.~\cite{Gupta:2021ahb} and reviewed in Sec.~\ref{sec:ChPT} to include the $\Delta(1232)$ as an explicit degree of freedom in Sec.~\ref{sec:Delta}. The change on including the $\Delta$ is small and does not change the conclusions, which are summarized in Sec.~\ref{sec:conclusions}.

\section{Lattice data}
\label{sec:lattice}

The lattice calculation presented in Ref.~\cite{Gupta:2021ahb} constructed Euclidean correlation functions using Wilson-clover fermions on six 2+1+1-flavor
ensembles generated using the highly improved staggered quark
action~\cite{Follana:2006rc} by the MILC
collaboration~\cite{Bazavov:2012xda}. These ensembles include data at $\mpi \approx 315$, $230$, and $138\MeV$. 
 To obtain the flavor-diagonal charges $g_S^q$, both connected and disconnected diagrams were calculated using the methodology presented in Refs.~\cite{Bhattacharya:2015wna,Gupta:2018qil}. Simultaneous fits to the zero momentum nucleon two-point, $C^{2\text{pt}}$, and three-point,
$C^{3\text{pt}}$, functions were made using their spectral
decomposition
\begin{align}
  C^{2\text{pt}}(\tau; \kk) &= \sum_{i=0}^3 |\mathcal{A}_i(\kk)|^2 e^{-M_i \tau},  \nonumber \\
C_\mathcal{S}^{3\text{pt}}(\tau; t) &=
   \sum_{i,j=0}^2 {\mathcal{A}_i} {\mathcal{A}_j^\ast}\langle i|\mathcal{S}|j\rangle e^{-M_i t - M_j(\tau-t)\
},
\label{eq:2pt3pt}
\end{align}
keeping four and three states, respectively. Here, $\tau$ denotes the source--sink separation and $t$ the time of the operator insertion, while
${\mathcal{A}}_i$ are the amplitudes for the
creation or annihilation of states by the nucleon interpolating operator.

The important observation was that current lattice data are not precise enough to resolve the excited-state masses $M_1$ and $M_2$. Fits using two strategies,   $\{4,3^\ast\}$ (standard fit, wide priors on the $M_{i>0}$) and
$\{4^{N\pi},3^\ast\}$ (excited-state fit, motivated by $\chi$PT, with narrow prior for $M_1$ centered around
the noninteracting energy of the almost-degenerate lowest positive-parity multihadron
states, $N(\mathbf{1}) \pi(- \mathbf{1})$ or
$N(\mathbf{0}) \pi(\mathbf{0}) \pi( \mathbf{ 0})$) gave similarly good 
fits, but vastly different results for $\spiN$. While the standard fit reproduces values around $40\MeV$, imposing multihadron ESC as in the $\{4^{N\pi},3^\ast\}$ fit gave $\approx 60$~MeV, thus removing the tension with the phenomenological value. The calculation needs validation, e.g., the conclusion is mainly driven by the single physical pion mass ensemble, however, it was supported by an analysis of ESC in $\chi$PT, as we will delineate in the following.  

\begin{figure*}[th]
\begin{center}
\includegraphics[width=0.75\textwidth]{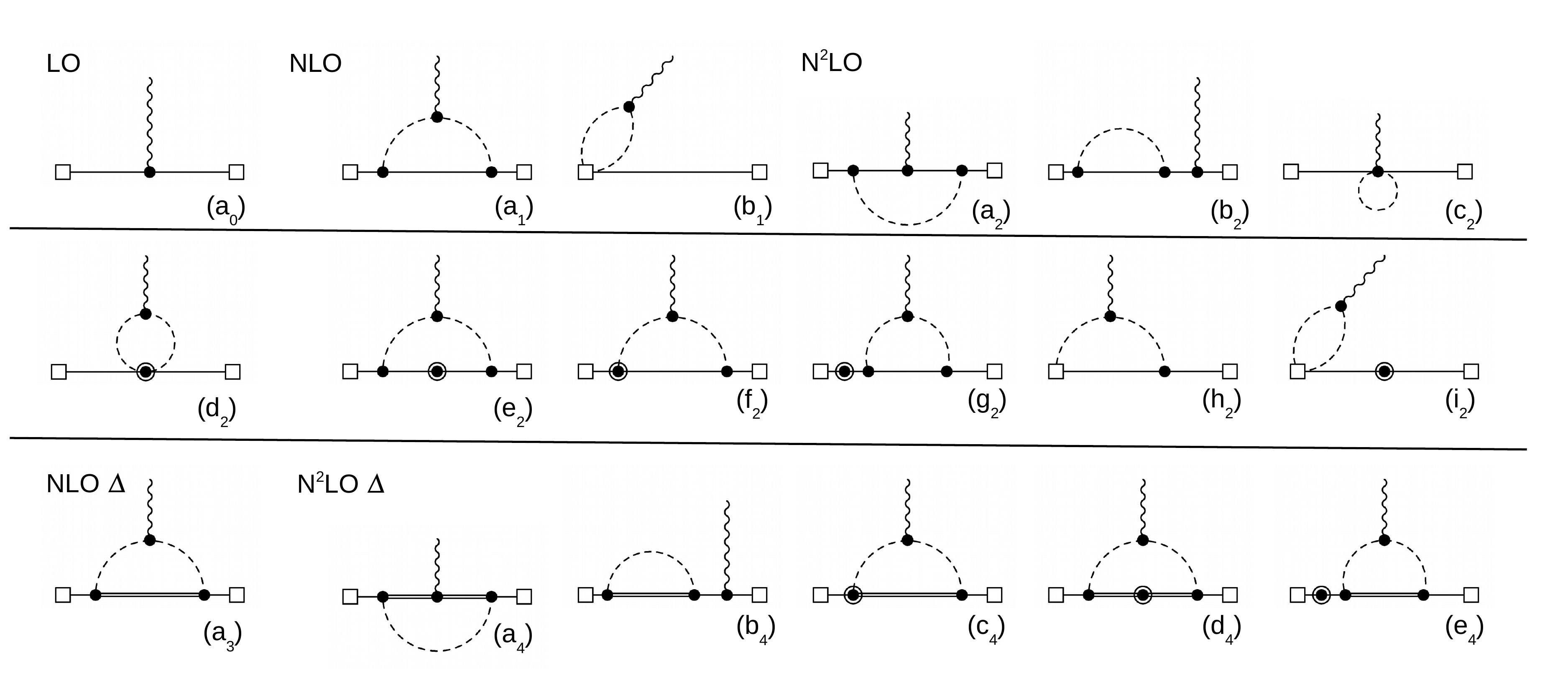}
\end{center}
\caption{Corrections to the scalar charge in $\chi$PT. Empty and full squares denote, respectively, insertions of the LO and NLO expansion of the source fields $\mathcal N$ and $\bar{\mathcal N}$. 
Plain, dashed, and wavy lines denote, nucleons, pions, and an insertion of the scalar source. Dots and circled dots denote LO and NLO vertices in the chiral Lagrangian.
Diagrams $(h_2)$ and $(i_2)$
are representative of  
N$^2$LO corrections arising from the chiral expansion of $\mathcal N$, which only produces negligible N$^2$LO recoil corrections.
The diagrams in the last row show the corrections induced by the $\Delta$ baryon, at NLO (diagram $(a_3)$) and N$^2$LO
(diagrams $(a_4)$ to $(e_4)$).
  \label{fig:chipt}}
\end{figure*}

\section{Excited states in chiral perturbation theory}
\label{sec:ChPT}

ESC has been studied before using $\chi$PT methods~\cite{Bar:2015zwa,Tiburzi:2015tta,Bar:2016uoj}. Given the subtle chiral expansion of $\spiN$, with chiral loops only suppressed by a single order compared to the tree-level contribution and even subleading loops enhanced due to the presence of the $\Delta(1232)$ as reflected by large values of the corresponding low-energy constants (LECs) $c_i$, we carried out a full next-to-next-to-leading-order (N$^2$LO) analysis, including the diagrams shown in Fig.~\ref{fig:chipt}. Expressed in terms of ratios of correlation functions $\mathcal R_S(\tau, t)$, which for $t,\tau\to\infty$ yield $\spiN$, we find
\begin{align}
 \mathcal R^{(1)}(\tau, t) &= \frac{3 g^2_A \mpi^2 }{8 F_\pi^2  L^3} \sum_{\kk}  \frac{\kk^2}{E_\pi^4}
 \bigg[ 1 -  
  e^{-E_{N \pi} t}  - e^{- E_{N \pi} t_B } + \frac{1}{2}  e^{ - E_{N \pi} \tau  } 
 + \frac{1}{4}e^{- 2 E_\pi t }  + \frac{1}{4} e^{- 2 E_\pi t_B } 
   \bigg] \nonumber \\
&      - \frac{3 \mpi^2 }{32 F_\pi^2} \frac{1}{L^3} \sum_{\kk}  \frac{1}{E_\pi^2}
    \left(e^{-2 E_\pi t } + e^{-2 E_\pi t_B}\right),\notag\\
 \mathcal R^{(2)}_{c_i}(\tau, t) &= -   \frac{3 \mpi^2}{4 F_\pi^2}
 \frac{1}{  L^3} \sum_{\kk}  \frac{ 1  }{E_\pi^3} 
\Big(  (c_2 + 2 c_3) E_\pi^2 + (2 c_1 -c_3) \mpi^2\Big)
 \bigg[ 1 -  
 \frac{1}{2} e^{-2 E_\pi  t} - \frac{1}{2}e^{-2 E_\pi t_B }  
  \bigg] \notag\\
  &+ \frac{3 \mpi^2}{ F_\pi^2}
 \frac{1}{  L^3} \sum_{\kk}  \frac{ 1  }{E_\pi} c_1,
 \label{R}
\end{align}
for the next-to-leading-order (NLO) result $\mathcal R^{(1)}$ and the by far most sizable N$^2$LO correction $\mathcal R^{(2)}_{c_i}$ involving the LECs $c_i$. The notation for the energies that appear in the sum over discrete momenta $\kk = 2 \pi \nn/L$ is
 $E_\pi = \sqrt{\kk^2 + \mpi^2}$, $\widetilde E_N = \sqrt{\mN^2 + \kk^2} - \mN$, $E_{N \pi}=E_\pi+\tilde E_N$, $\mN$ and $\mpi$ are the full nucleon and pion mass at the corresponding quark mass in the lattice simulation,
and the time difference $ \tau - t$ is denoted by $t_B$. Note that $E_{N\pi}$   subsumes some of the N$^2$LO recoil corrections. For the remaining contributions as well as finite-volume corrections, evaluated as the difference of the ground-state contribution in Eq.~\eqref{R} to the continuum result~\cite{Beane:2004tw}, we refer to Ref.~\cite{Gupta:2021ahb}.

Using the values for the $c_i$ from Refs.~\cite{Hoferichter:2015tha,Hoferichter:2015hva}, we find that both contributions in Eq.~\eqref{R} produce large, negative corrections, that make  $\spiN$ too small if these ESC are not taken into account. Depending on the details of the lattice ensemble in question, we find that NLO and N$^2$LO contributions can each generate up to $-10\MeV$ at $t=\tau/2\sim (0.5\text{--}0.7)\rm fm$. The combined effect is to reduce the $\approx 60$~MeV value on the physical mass ensemble to $\approx 40$~MeV. Previous calculations using the direct method did not include these multihadron states in their analysis~\cite{Aoki:2021kgd}, thus creating the tension between direct lattice calculations of $\spiN$ and phenomenology. A similar argument about the relevance of these multihadron ESC also applies to the Feynman--Hellmann method. A standard worry, however, in such $\chi$PT analyses is that the results need to be applied at scales at which the convergence of the heavy-baryon expansion is not guaranteed. Here we extend the analysis by considering the effect of including the $\Delta(1232)$ as an explicit degree of freedom to gauge the stability of the chiral expansion. These new results are presented in the next section.

\section{Including the \texorpdfstring{$\boldsymbol{\Delta(1232)}$}{\textDelta(1232)} resonance as explicit degree of freedom}
\label{sec:Delta}

The first contribution from the $\Delta(1232)$ arises at NLO and is shown in diagram $(a_3)$ in Fig. \ref{fig:chipt}. We find
\begin{align}
\mathcal R^{(1)}_{\Delta}(\tau, t) 
 &=\frac{h_A^2\mpi^2}{6F_\pi^2}  \left(1 - \frac{\epsilon}{3}\right) \frac{1}{L^3} \sum_{\kk}\frac{\kk^2}{E_\pi^{3}(E_\pi + \Delta)^2}
 \Bigg\{  2(2 E_\pi + \Delta)-(3E_\pi + \Delta) \left( e^{-2 E_\pi t} + e^{- 2 E_\pi t_B}\right)\notag\\
&+2 E_\pi
e^{- (E_\pi + \Delta) (t+ t_B)}+4E_\pi^2\bigg(e^{-E_\pi t}\frac{e^{-E_\pi t}-e^{-\Delta t}}{E_\pi-\Delta}+e^{-E_\pi t_B}\frac{e^{-E_\pi t_B}-e^{-\Delta t_B}}{E_\pi-\Delta}\bigg)\Bigg\},
\label{DeltaLO}
 \end{align}
 written in a form that makes the cancellation of the singularities at $E_\pi=\Delta=M_\Delta-\mN$ apparent.
 At NLO, the $\Delta$--nucleon mass splitting in Eq.  \eqref{DeltaLO} should be interpreted strictly in the chiral limit, $\Delta= \Delta^{(0)} =M^{(0)}_\Delta-\mN^{(0)}$. Here $\epsilon=(4-d)/2$  is the regulator in dimensional regularization, as needed to reproduce the continuum result from Eq.~\eqref{DeltaLO}, and $h_A$ denotes the $\pi N\Delta$ coupling in the conventions of Ref.~\cite{Siemens:2016jwj}. The finite-volume corrections can again be obtained by comparing the ground-state contribution to the continuum, i.e., momentum sums versus integrals, leading to
\begin{align}
    \Delta_L^{(1), \Delta}\spiN=\frac{h_A^2 \mpi^2}{6\pi^2F_\pi^2}\sum_{\nn \neq \boldsymbol{0} }\int_0^\infty d\lambda \bigg[3 K_0(L \sqrt{\beta} |\nn|)-\sqrt{\beta}L|\nn|K_1(L \sqrt{\beta} |\nn|)\bigg],
\end{align}
with $\beta=\lambda^2+2\lambda\Delta+\mpi^2$ and Bessel functions $K_0$, $K_1$.

N$^2$LO corrections arise from recoil corrections to the $\Delta$ propagator and to the $\Delta$--nucleon vertices, as well as from the LEC $c_1^{\Delta}$. The latter contributes in two ways, by 
mediating the coupling of the scalar charge to the $\Delta$ baryon and by providing a quark-mass dependent correction to the $\Delta$ mass. 
Diagrams $(d_4)$ and $(e_4)$ can be absorbed by shifting 
\begin{align}
\Delta\to \tilde E_{\Delta}&\equiv\sqrt{\left(\mN^{(0)} + \Delta^{(0)} - 4 \mpi^2 c_1^\Delta  \right)^2 + \kk^2} - \left(\mN^{(0)} - 4 \mpi^2 c_1 \right) \notag\\
&= \Delta^{(0)} 
 - 4 \mpi^2 ( c_1^\Delta - c_1 ) + \frac{\kk^2}{2 \mN^{(0)}} + \mathcal{O}\Big(\mN^{-2}\Big),
 \label{Delta_recoil}
\end{align}
in the NLO  contribution~\eqref{DeltaLO}.
$c_1$ and $c_1^\Delta$ quantify pure explicit-symmetry-breaking terms, in such a way that the  nucleon and $\Delta$ becoming degenerate in the large-$N_c$ limit strongly suggesting $c_1=c_1^\Delta$ up to $1/N_c$ corrections~\cite{Siemens:2016jwj}.
Effectively, we capture these corrections by using the physical values of the nucleon and $\Delta$ masses in
$\tilde E_{\Delta}$ and in Eq. \eqref{DeltaLO}.

The remaining N$^2$LO corrections are given by diagrams
$(a_4)$, $(b_4)$, and $(c_4)$ and by analogous corrections to the two-point function, which yield
\begin{align}\label{Rc1Delta}
 \mathcal R^{(2)}_{\Delta}(\tau, t) 
 &=\frac{2h_A^2\mpi^2}{3F_\pi^2}\left(1 - \frac{\epsilon}{3}\right) 
 \left( \frac{1}{\mN} - 4  \left(c_1^\Delta - c_1\right) \right) \notag \\
&  \times  \frac{1}{L^3} \sum_{\kk} \frac{\kk^2}{E_\pi (E_\pi + \Delta)^2} \bigg[1 - e^{- (E_\pi+ \Delta) t} - e^{- ( E_\pi + \Delta) t_B} + e^{- (E_\pi+ \Delta) (t+t_B)} \bigg] \notag\\
& +  \frac{2h_A^2\mpi^2}{3F_\pi^2\mN} \left(1 - \frac{\epsilon}{3}\right)
 \frac{1}{  L^3} \sum_{\kk}  \frac{ 1  }{E_\pi} 
 \bigg[ 1 -  
 \frac{1}{2} e^{-2 E_\pi  t} - \frac{1}{2}e^{-2 E_\pi t_B }  
  \bigg]. 
\end{align}
To obtain these expressions, we have chosen a renormalization scheme that reproduces the continuum results from Ref.~\cite{Siemens:2020vop}, ensuring consistency with the LECs from Ref.~\cite{Siemens:2016jwj}. 
The last line of Eq. \eqref{Rc1Delta} leads to a shift in the couplings $c_2+2c_3$ in Eq.~\eqref{R} by $\Delta(c_2+2c_3)=-8h_A^2/(9\mN)$.
The first line contains a recoil correction and 
a correction proportional to $c_1^{\Delta} - c_1$. As discussed above, the latter is expected to vanish in the large-$N_c$ limit.

\begin{figure}
    \centering
\includegraphics[width=0.495\textwidth]{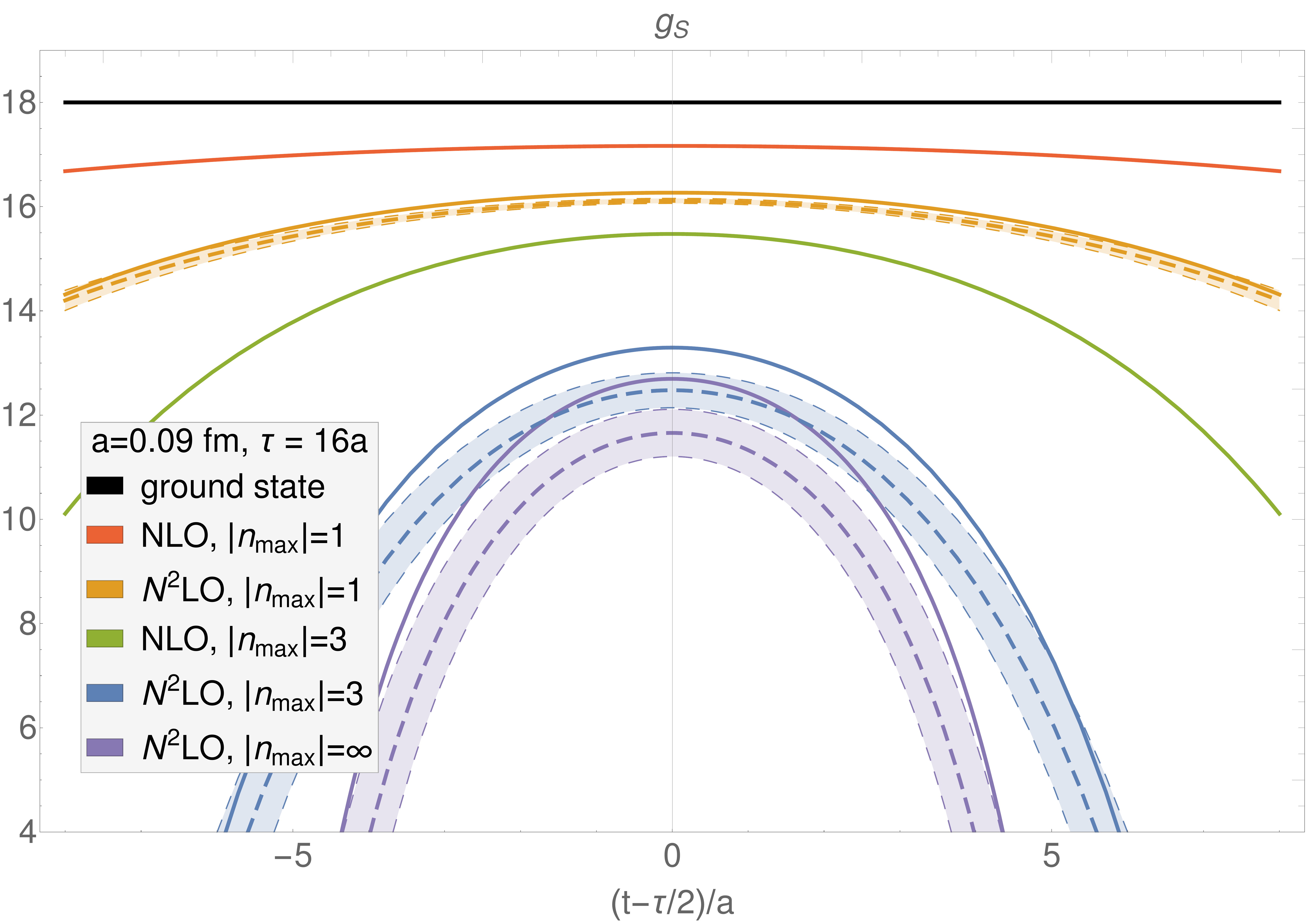}
    \includegraphics[width=0.495\textwidth]{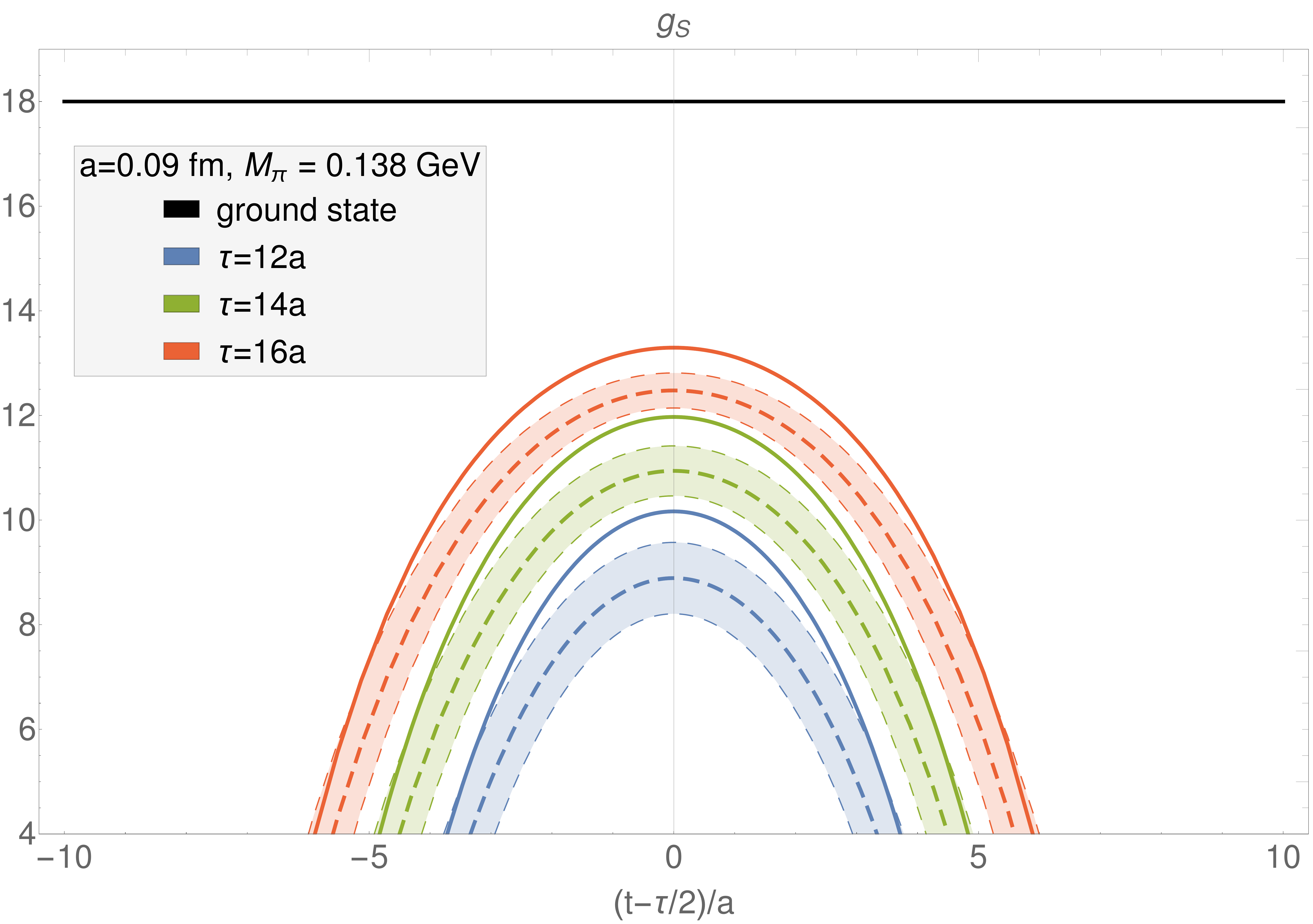}
    \caption{
    (Left) ESC from different truncations to the isoscalar scalar charge $g_S$ in $\chi$PT. (Right) Estimates
for $R_S(\tau, t)$ from the N$^2$LO analysis for the a09m130 ensemble. In both cases, the dashed bands indicate the outcome of the full N$^2$LO analysis including the $\Delta(1232)$, in comparison to the $\Delta$-less results using solid lines. The figure generalizes Fig.~6 in Ref.~\cite{Gupta:2021ahb}.
    }
    \label{fig:Delta}
\end{figure}

Equation~\eqref{DeltaLO} is evaluated numerically with the resummed shift~\eqref{Delta_recoil} and the physical value of the $\Delta$--nucleon mass splitting. 
Further, we vary $c_1-c_1^\Delta$ between $\pm |c_1|/N_c$ as an estimate of the corresponding uncertainty, leading to the bands in Fig.~\ref{fig:Delta} for the full N$^2$LO analysis including the $\Delta(1232)$ baryon. For comparison, the results from the $\Delta$-less calculation are shown by solid lines, both for different truncations in the chiral order and sum over momenta $\kk$ (left) and source--sink separations (right). In particular, the figure illustrates that the corrections beyond the N$^2$LO $\Delta$-less results are small, much smaller than the shift between NLO and N$^2$LO results. These findings indicate that the chiral expansion is reasonably stable, with the main effects indeed captured by the leading-loop contributions and the $\Delta$-enhanced corrections from the $c_i$ that were already included in Ref.~\cite{Gupta:2021ahb}.

\section{Conclusions}
\label{sec:conclusions}

We have summarized the main arguments from Ref.~\cite{Gupta:2021ahb} that provide a resolution of the tension between phenomenological determinations of $\spiN$ and direct lattice calculations. We demonstrated the impact of ESC in both the lattice calculation and in $\chi$PT  up to N$^2$LO. Here we have extended the $\chi$PT calculation to include the $\Delta(1232)$ resonance as an explicit degree of freedom to assess the stability of the chiral expansion, and find that for $\spiN$ the impact is remarkably small, thereby corroborating the conclusions from the $\Delta$-less heavy-baryon analysis presented in Ref.~\cite{Gupta:2021ahb}.

\section*{Acknowledgments}
M.H.\ acknowledges financial support by the Swiss National Science Foundation (Project No.\ PCEFP2\_181117). T.B.\ and R.G.\ were partly supported by the DOE HEP under Award No.\ DE-AC52-06NA25396. T.B., R.G., E.M., S.P., and B.Y.\ were partly supported by the LANL LDRD program.

\bibliographystyle{JHEP}
\bibliography{ref}

\end{document}